# True Atomic-Resolution Imaging under Ambient Conditions via Conductive Atomic Force Microscopy

Saima A. Sumaiya[1] and Mehmet Z. Baykara[1,a]

[1]Department of Mechanical Engineering, University of California Merced, Merced, California 95343, USA

Atomic-scale characteristics of surfaces dictate the principles governing numerous scientific phenomena ranging from catalysis to friction. Despite this fact, our ability to visualize and alter surfaces on the atomic scale is severely hampered by the strict conditions under which the related methods are operated to achieve high spatial resolution. In particular, the two prominent methods utilized to achieve atomic-resolution imaging – scanning tunneling microscopy (STM) and noncontact atomic force microscopy (NC-AFM) – are typically performed under ultrahigh vacuum (UHV) and often at low temperatures. Perhaps more importantly, results obtained under such well-controlled, pristine conditions bear little relevance for the great majority of processes and applications that often occur under ambient conditions. As such, a method that can robustly image surfaces on the atomic scale under ambient conditions has long been thought of as a "holy grail" of surface science. Here, by way of a proof-of-principle measurement on molybdenum disulfide ($MoS_2$), we report that the method of conductive atomic force microscopy (C-AFM) can be utilized to achieve *true* atomic-resolution imaging under ambient conditions as proven by the imaging of a single atomic vacancy, without any control over the operational environment or elaborate sample preparation. While the physical mechanisms behind this remarkable observation are not elucidated yet, our approach overcomes many of the classical limitations associated with STM and NC-AFM, and the findings herald the potential emergence of C-AFM as a powerful tool for atomic-resolution imaging under ambient conditions.

---

[a] Author to whom correspondence should be addressed: mehmet.baykara@ucmerced.edu

**Introduction**

Atomic-resolution imaging provides critical information about the structure and properties of material surfaces, with implications for a multitude of fields in science and technology, ranging from the design of small-scale electronic devices, all the way to surface reactivity and heterogenous catalysis. Atomic-resolution imaging also has the potential to provide feedback for bottom-up synthesis of materials with predictable properties. The first direct, real-space atomic-resolution imaging of a surface was made possible thanks to the invention of the scanning tunneling microscope (STM) [1]. Despite its groundbreaking success, a major limitation associated with STM is that the topographic and electronic information are convoluted in the tunnelling current signal, which often makes the physical interpretation of recorded images complicated on the atomic scale. Another prominent member of the scanning probe microscopy (SPM) family, the atomic force microscope (AFM) [2], while very versatile in terms of the information it can collect and the range of material surfaces it can be applied to, is unable to achieve atomic resolution in the widely utilized "contact mode". This is due to the fact that the repulsive contact formed between the probe tip and the sample surface leads to contact areas that are at least several atoms wide, resulting in an averaging of force interactions across this rather blunt contact and consequently the inability to image individual atomic vacancies and thus, to achieve *true* atomic resolution [3-4]. On the other hand, nearly a decade after the invention of the AFM, a method named noncontact atomic force microscopy (NC-AFM) was introduced by Giessibl and then Morita's group, with which atomic-resolution imaging can be achieved on a wide range of sample surfaces by means of detecting the short-range interaction forces between the sharp probe tip and the sample surface [5-6].

A critical limitation associated with STM and NC-AFM is that they typically cannot be performed under ambient conditions toward atomic-resolution imaging (although atomic/molecular-resolution imaging has been achieved on certain samples under liquid environment by NC-AFM [7], and there has been one demonstration of true atomic resolution imaging via NC-AFM on a calcite surface in air [8]). When exposed to ambient conditions, most surfaces are covered with a layer of contaminants adsorbed from the environment. In order to suppress the adsorption of contaminants and record data that are representative of actual sample surface, STM and NC-AFM imaging is typically performed under ultrahigh vacuum (UHV) conditions. In addition, in the case of NC-AFM, measuring interaction forces between the probe tip and the sample via tracking the oscillation frequency of the cantilever with a good signal-to-noise ratio typically requires slow scanning, which in turn may require low-temperature operation or elaborate correction algorithms to suppress the effects of thermal drift [9]. It is also worth noting that mastering the skill to reliably operate multiple simultaneous feedback loops associated with NC-AFM takes years to develop. Furthermore, operation under UHV and often, low temperatures, are necessitated to achieve stable operation in STM and NC-AFM, whereby even minute changes in tip structure or chemistry, which may be induced by thermally activated events and the presence of contaminants, may lead to issues of irreproducibility in the obtained results, leading to a severe lack of experimental robustness. Overall, the requirement for complex equipment makes the atomic-resolution STM and NC-AFM techniques expensive and thereby, exclusive. Even more importantly, properties measured and processes observed under the ideal yet unrealistic UHV environment bear little relevance for technological applications such as heterogenous catalysis which often take place at elevated pressures, leading to the long-standing issue of the "pressure gap" in surface science [10]. Similar arguments can be made, *e.g.* for measuring electronic properties of surfaces under UHV conditions, with limited relevance to device applications that will more than likely operate under ambient conditions. Motivated by all of these arguments, a technique that can reliably and reproducibly provide true atomic-resolution images of a wide range of sample surfaces under ambient atmosphere and at room temperature would be ideal, and could perhaps even be considered a "holy grail" of surface science.

Another SPM technique that has been explored to achieve atomic-resolution is conductive atomic force microscopy (C-AFM), where a conductive probe tip scans the sample surface in contact mode under the application of a bias. Early attempts resulted in "lattice resolution" imaging under UHV conditions on the prototypical sample surface of highly oriented pyrolytic graphite (HOPG) [11-12], whereby features corresponding to the lattice spacing of HOPG are imaged but no individual carbon atoms or atomic defects. More recently, under a controlled nitrogen environment, Bampoulis *et. al.* imaged $WSe_2$ with atomic resolution [13], while Nowakowski *et. al.* also reported atomic-resolution imaging via C-AFM under a nitrogen environment, by way of visualizing atomic defects on $WS_2$ [14]. While these efforts demonstrate the high spatial resolution capabilities of the C-AFM method on specific samples, the restriction to the clean yet unrealistic nitrogen environment has similar implications as UHV, namely the requirement of additional experimental equipment, and the limited relevance of the obtained results for realistic applications.

Motivated by the arguments above, we demonstrate here that the method of C-AFM can be utilized to achieve true-atomic resolution imaging under ambient conditions by means of a proof-of-principle measurement on the prototypical sample of molybdenum disulfide ($MoS_2$). In particular, the combined demonstration of the simultaneous imaging of three types of atomic sites and a single atomic defect herald the emergence of C-AFM as a powerful tool for atomic-resolution imaging under ambient conditions.

**Methods**

As a substrate for MoS$_2$, we used Si/SiO$_2$ chips coated with a ~2 nm thick adhesion layer of Ti and a ~50 nm thick Au film. MoS$_2$ flakes were mechanically exfoliated on top of the Au-coated substrate via the scotch tape method [15]. Finally, silver paint was applied on one side, bridging the conductive specimen holder and the Au film on which the MoS$_2$ flakes are located. The C-AFM measurements were performed using a commercial AFM (*Asylum Research, Cypher VRS*) under ambient conditions (Temperature: 22–23 °C; Relative humidity: 20–50%). Samples were inserted inside the AFM chamber without any prior treatment for surface cleaning. Commercially available, doped-diamond-coated conductive tips (*NanoSensors, CDT-CONTR*) were used for imaging. Initially, contact mode imaging on large areas (*e.g.* 10 × 10 μm) was performed without any bias to find regions of interest on the sample surface. For atomic-resolution imaging, small areas such as 5 × 5 nm were scanned in contact mode with an applied bias. Imaging was usually performed under only the adhesion force acting between the tip and the sample, with no normal load applied in addition. The magnitude of the adhesion force between the tip and sample varied between 20 to 30 nN. The bias voltage was always applied to the sample. An additional resistance of 10 MΩ was used in series occasionally to limit the current. The average periodicities of the atomic features observed in the images were calculated using the software *Gwyddion* by way of Fast Fourier Transform (FFT).

**Results and Discussion**

As a benchmark test, we first employed our methodology on HOPG and obtained similar results to what is reported in the literature [12,16], *i.e.* periodic features separated by ~2.50 Å which does not correspond to the interatomic distance of 1.42 Å that is expected on an HOPG (0001) surface. Combined with the fact that we do not see any missing atoms or other types of defects in any of our images, what we observe can be only classified as "lattice resolution" and true atomic-resolution cannot be claimed.

The key signature of true atomic-resolution imaging is the capability to identify single atomic defects such as individual vacancies. In order to further explore the possibility of true atomic-resolution imaging capabilities on a material system that is expected to feature a higher density of defects than HOPG, we switched our attention to another prototypical sample for nanoscience: $MoS_2$.

The results obtained on $MoS_2$ via C-AFM imaging are in striking difference to HOPG. As demonstrated by the representative image shown in Fig. 1A, we are able to achieve ultimate spatial resolution on $MoS_2$ under ambient conditions, signified by the fact that we are able to simultaneously resolve three types of atomic sites (see green, purple, and yellow circles in Fig. 1A) with the expected periodicity of ~3.2 Å [17] that feature high, low, and intermediate current, which in turn would correspond, in no particular order, to the three symmetry sites associated with the $MoS_2$ (001) surface: Mo atoms, S atoms, and hollow sites at the center of the hexagons formed by the two types of atoms. In addition to the fact that we are able to resolve all three atomic sites in our image, what's even more important is the fact that we can clearly identify a single atomic defect in our image, manifesting as a missing "bright spot" in Fig. 1A that is indicated by the white arrow. This latter observation unequivocally proves that we are able to achieve true atomic-resolution imaging via the method of C-AFM under ambient conditions. Comparing our results to what has been achieved on $MoS_2$ via STM under UHV conditions, we tentatively assign the single atomic defect in our image to an S vacancy, due to the rather isolated effect it has on the local density of states, in contrast to electronic effects extending over several to several tens of nanometers for Mo vacancies and various dopants [18-19].

To take the first step toward understanding the physical mechanisms involved in our ability to achieve true atomic-resolution imaging under ambient conditions while the tip apex is *in contact* with the sample surface, we present, in Fig. 1B, the topography map that is simultaneously recorded with the C-AFM (*i.e.* current) map in Fig. 1A. There are well-defined differences between the two channels. In particular, we observe a vague periodicity in the topography image consisting of stripes of ~0.5 Å height, reminiscent of stick-slip behavior [20]. The fact that we cannot resolve clearly defined atomic sites in the topography image already indicates that the mechanism that results in atomic-resolution in the C-AFM image is not the same as the one responsible for atomic-scale stick-slip patterns in the topography image. This is in accordance with the idea that "lattice resolution" maps recorded in contact-mode AFM typically arise from an averaged behavior commonly captured by recording tip-sample interactions across a contact that consists of multiple atoms, and thus is not atomically sharp. Consequently, it is not surprising that we are unable to resolve the single atomic defect in the topography channel, as shown in Fig. 1B.

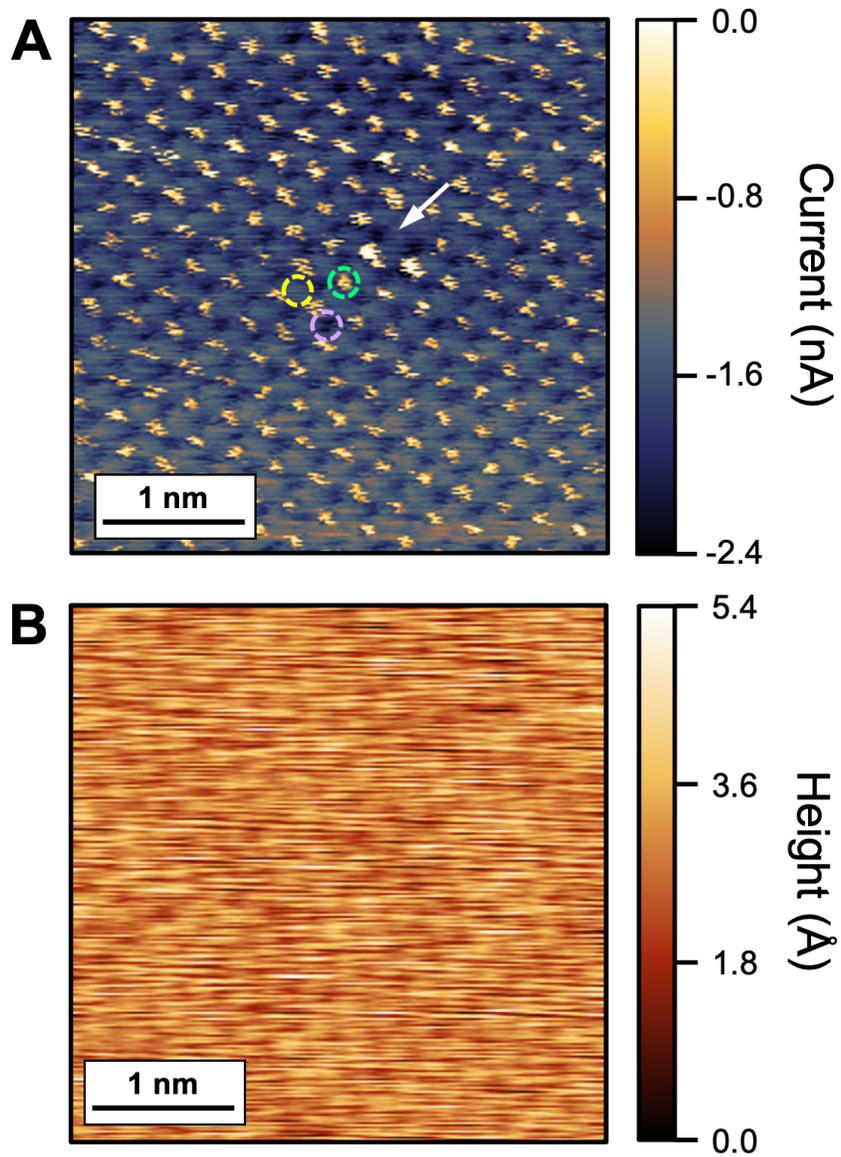

**Figure 1. True atomic-resolution imaging under ambient conditions via C-AFM.** (**A**) C-AFM image obtained on MoS$_2$ exhibiting three types of atomic sites characterized by low, high, and intermediate current (see corresponding green, purple, and yellow circles, respectively). This image clearly captures a single atomic defect as indicated by the white arrow. (**B**) Simultaneously obtained topography image showing an atomic-scale stick-slip pattern, with no trace of the single atomic defect detected in part (A) (bias voltage: -1.2 V, applied normal load: 0.0 nN).

## Conclusions

While the findings reported here clearly demonstrate the potential of C-AFM as a revolutionary tool for atomic-resolution surface science under ambient conditions, much work needs to be done to understand the mechanisms that enable this rather surprising capability, to define the operational parameter space under which atomic-resolution can be retained, explore how far the realm of applicability can be expanded in terms of sample materials, and analyze applications beyond atomic-resolution imaging and into the domain of structural as well as electronic manipulation.

## Acknowledgements

This work was supported by the Air Force Office of Scientific Research (AFOSR) Award No. FA9550-19-1-0035.

## References

[1] G. Binnig, H. Rohrer, C. Gerber, and E. Weibel, Phys. Rev. Lett. **50**, 120 (1983).
[2] G. Binnig, C. F. Quate, and C. Gerber, Phys. Rev. Lett. **56**, 930 (1986).
[3] P. Eaton and P. West, *Atomic Force Microscopy*, Oxford University Press (2010).
[4] M. Z. Baykara and U. D. Schwarz, *Atomic Force Microscopy: Methods and Applications*, in Encyclopedia of Spectroscopy and Spectrometry, 3rd ed., Elsevier (2017).
[5] F. J. Giessibl, Science **267**, 68 (1995).
[6] H. Ueyama, M. Ohta, Y. Sugawara, and S. Morita, Jpn. J. Appl. Phys. **34**, L1086 (1995).
[7] T. Fukuma and R. Garcia, ACS Nano **12**, 11785 (2018).
[8] D. S. Wastl, M. Judmann, A. J. Weymouth, and F. J. Giessibl, ACS Nano **9**, 3858 (2015).
[9] M. Z. Baykara, *Noncontact Atomic Force Microscopy for Atomic-scale Characterization of Material Surface*s, in Surface Science Tools for Nanomaterials Characterization, Springer (2015).
[10] R. Imbihl, R. J. Behm, and R. Schlögl, Phys. Chem. Chem. Phys. **9**, 3459 (2007).
[11] D. P. E. Smith, G. Binnig, and C. F. Quate, Appl. Phys. Lett. **49**, 1166 (1986).
[12] M. Enachescu, D. Schleef, D. F. Ogletree, and M. Salmeron, Phys. Rev. B **60**, 16913 (1999).
[13] P. Bampoulis, K. Sotthewes, M. H. Siekman, and H. J. W. Zandvliet, ACS Appl. Mater. Interfaces **10**, 13218 (2018).
[14] K. Nowakowski, H. J. W. Zandvliet, and P. Bampoulis, Nano Lett. **19**, 1190 (2018).
[15] K. S. Novoselov, A. K. Geim, S. V. Morozov, D. Jiang, Y. Zhang, S. V. Dubonos, I.V. Grigorieva, and A. A. Firsov, Science **306**, 666 (2004).
[16] C. Rodenbücher, G. Bihlmayer, W. Speier, J. Kubacki, M. Wojtyniak, M. Rogala, D.

Wrana, F. Krok, and K. Szot, Nanoscale **10**, 11498 (2018).
[17] A. Yan, C. S. Ong, D. Y. Qiu, C. Ophus, J. Ciston, C. Merino, S. G. Louie, A. Zettl, J. Phys. Chem. C **121**, 22559 (2017).
[18] R. Addou, L. Colombo, and R. M. Wallace, ACS Appl. Mater. Interfaces **7**, 11921 (2015).
[19] P. Bampoulis, R. van Bremen, Q. Yao, B. Poelsema, H. J. W. Zandvliet, and K. Sotthewes, ACS Appl. Mater. Interfaces **9**, 19278 (2017).
[20] H. Hölscher, U. D. Schwarz, O. Zwörner, and R. Wiesendanger, Phys. Rev. B, **57**, 2477 (1998).